\documentclass[11pt]{article}

\usepackage{verbatim}
\usepackage{amsmath}
\usepackage{amsfonts}
\usepackage{amssymb}
\usepackage{graphicx}
\usepackage{bm}
\usepackage{color}
\usepackage{mathrsfs}
\usepackage{epsf}
\usepackage[hypertex]{hyperref}

\begin{document}

\newcommand{\la}{\label}
\newcommand{\be}{\begin{eqnarray}}
\newcommand{\ee}{\end{eqnarray}}
\newcommand{\ls}{\!\prec\!\!}
\newcommand{\rs}{\!\!\succ\;}
\newcommand{\mb}{\mathbf}
\newcommand{\der}{\partial}
\newcommand{\p}{\partial}
\newcommand{\w}{\tilde}
\newcommand{\sgn}{\,{\mathrm{sgn}}}
\newcommand{\tr}{\,{\mathrm{tr}}\,}
\newcommand{\re}{{\mathrm{Re}}\,}
\newcommand{\im}{{\mathrm{Im}}\,}
\newcommand{\cO}{\mathcal{O}}

\title{Statistics of harmonic measure and winding of critical curves
from conformal field theory}

\author{A. Belikov$^1$, I. A. Gruzberg$^1$, I. Rushkin$^2$ \\
$^1$James Franck Institute, University of Chicago, Chicago IL
60637
\\
$^2$School of Physics, Georgia Institute of Technology,
Atlanta, GA 30332}

\date{May 29, 2008}

\maketitle

\begin{abstract}

Fractal geometry of random curves appearing in the scaling
limit of critical two-dimensional statistical systems is
characterized by their harmonic measure and winding angle. The
former is the measure of the jaggedness of the curves while the
latter quantifies their tendency to form logarithmic spirals.
We show how these characteristics are related to local
operators of conformal field theory and how they can be
computed using conformal invariance of critical systems with
central charge $c \leqslant 1$.

\end{abstract}

\section{Introduction}

Geometric properties of critical two-dimensional systems of
statistical mechanics have attracted considerable interest of
both physicists and mathematicians in recent years. On the one
hand, methods of quantum gravity have been successfully applied
by Duplantier to obtain the multifractal spectrum of harmonic
measure on critical cluster boundaries \cite{Duplantier-PRL,
Duplantier}. On the other hand, the invention of the
Schramm-Loewner evolution (SLE) by Schramm \cite{Schramm} has
made it possible to study conformally-invariant critical curves
rigorously (several reviews of SLE are available by now; see,
for example, Refs. \cite{Lawler-book, Kager-Nienhuis, Cardy,
Bauer-Bernard-big-review, my review}).

Conformal field theory (CFT ) has been a natural and
traditional language for description of critical statistical
systems in two dimensions. It is important to understand how
fractal properties of critical curves can be obtained from CFT.
A part of the emerging picture is that the stochastic geometry
of critical curves can be studied by traditional methods of
CFT. As any local field theory, CFT focuses on correlation
functions of local operators. It turns out that the stochastic
geometry of critical curves is related to correlators of
certain primary fields in CFT. This connection, first discussed
in Ref. \cite{Bauer-Bernard-CMP}, has been further developed in
our papers \cite{BRGW-PRL, RBGW-long-paper} where we have
identified the curve-creating primary operators and considered
their correlators with other primaries serving as ``probes'' of
harmonic measure. Along this way we have reproduced
Duplantier's results for multifractal exponents associated with
harmonic measure.

In addition to harmonic measure, complicated planar domains and
their boundaries are characterized by winding or rotation. The
notion of rotation spectrum for planar domains has been
introduced by Binder in Ref. \cite{Binder-thesis}. Then the
mixed rotation and harmonic measure spectrum has been found
exactly for critical curves by Duplantier and Binder
\cite{Duplantier-Binder}. In that paper the authors combined an
earlier Coulomb gas approach to the distribution of winding
angles of critical curves \cite{DupSal} with the quantum
gravity methods of Refs. \cite{Duplantier-PRL, Duplantier}.

In this paper we extend our previous analysis \cite{BRGW-PRL,
RBGW-long-paper} to include the winding (rotation) of critical
curves. As it happens, to account for rotation of critical
curves, we need to consider CFT primary fields with {\it
complex} weights and charges. The procedure involves an
analytic continuation of chiral correlation functions and then
gluing the chiral sectors to obtain real but angular dependent
correlators. The paper is organized as follows. In Section
\ref{formulation of problem} we formulate the problem and give
heuristic definitions of various objects of interest including
harmonic measure and rotation of critical curves. In Section
\ref{one point} we consider a special case of a single probe of
harmonic measure and rotation near a star configuration of
several critical curves. Then we go on to the general case,
where several probes can be placed in the vicinity of a
critical star configuration. We conclude in Section
\ref{discussion} with a discussion and comparison with other
works.

\section{Description of the problem}
\label{formulation of problem}

Consider a two-dimensional critical statistical system. Such
systems can be formulated as critical points in an ensemble of
curves on a lattice, e.g. the loops of the $O(n)$ model or the
high-$T$ expansion loops of the $Q$-state Potts model. When the
system is critical, these curves are called critical curves. In
the continuum limit, statistical systems are invariant under
conformal transformations and can be described by a CFT with a
certain central charge $c$. In this case the critical curves
are fluctuating fractal curves.

Any conformal transformation in two dimensions locally looks
like a combination of rotation and a scale transformation
(dilatation). For a given composition of dilatation and
rotation in the plane there exist curves which remain
invariant. They are logarithmic spirals $\phi = \lambda\ln r$.
The quantity $\lambda$ characterizes the winding of a spiral
and is called the rate of rotation. For a pure dilatation the
spirals degenerate into straight lines (no rotation:
$\lambda=0$) which may intersect and form corners with
arbitrary opening angles.

Since any conformal transformation is locally a composition of
dilatation and rotation, a conformally invariant critical curve
can be thought of as an assembly of elementary corners and
logarithmic spirals at all scales. The so called multifractal
analysis aims at determining the fractal dimensions of the
subsets of a critical curve with given opening angle and
winding. These fractal dimensions form a continuous family of
multifractal exponents.

Harmonic measure is another way of characterizing the
complicated fractal geometry of a critical curve. In a simple
electrostatic analogy we can imagine that a critical cluster is
charged with a unit charge that all goes to the boundary of the
cluster. Charge density (which is the density of harmonic
measure) on the boundary is very uneven and lumpy, and can be
characterized by its moments. Therefore, in the ensemble of
critical curves it is natural to characterize their fractal
geometry by the statistics of local winding and harmonic
measure near some point on the curve.

Locally, near a charged corner the scaling of the charge
density or, equivalently, of the magnitude of the electric
field $E$ is determined by the opening angle $\beta$ at the
corner: $r E(r) \sim r^{\pi/\beta}$. What will be important for
us in the following is that this power-law behavior is the same
along all straight lines that converge at the apex of the
corner. All these lines intersect the equipotentials at
constant angles $\theta$, and only one of them is an electric
field line (corresponding to $\theta = \pi/2$). In mathematics
such lines are sometimes called ``slanted Green lines'', while
the electric field lines are called ``Green lines''. In a
general situation we will denote the slanted Green line that
goes from the origin and forms the angle $\theta$ with
equipotentials by $\gamma_\theta$.

Similar statements can be made for a logarithmic spiral. The
information about the rate of rotation $\lambda$ of a spiral is
contained in the winding of any of the slanted Green lines that
emanate from the spiral's origin. It is easy to see that the
line $\gamma_\theta$ is the rotated logarithmic spiral $\phi =
2\theta + \lambda \ln r$ (with $\theta = \pi/2$ again
corresponding to the unique electric field line going from the
origin), and we can follow any of them to ``measure''
$\lambda$.

The information on the scaling of the magnitude of the electric
field and the winding of a slanted Green line near a certain
point on the critical curve is contained in the value of the
derivative of a uniformizing conformal map $w(z)$ that maps the
exterior of the critical cluster onto a standard simple domain.
In these terms, the electrostatic potential of the curve is $-
\log |w(z)|$, the magnitude  of the field near the curve is $E
= |w'(z)|$, and the direction of the field is related to $\arg
w'(z)$. (In terms of the complex potential $\Phi(z) = - \log
w(z)$ we have $\Phi'(z) = - E_x + i E_y$.) We will use this
idea extensively in what follows. It is convenient for our
purposes to restrict the statistical ensemble so that a
critical curve always passes through a fixed point, and work in
the vicinity of this point.

In this paper we consider only dilute critical systems, which
means that critical curves generically do not intersect (see,
however, the discussion of exceptional star configurations
below). This choice is dictated by the nature of our problem.
In the dense phase where critical curves have multiple points,
the harmonic measure is only supported on the external
perimeters of critical clusters (think of electric charge
spreading on the surface of a conductor). The external
perimeters are always simple critical curves described by a
model in the dilute phase, and it is sufficient to consider
dilute systems in our problem. The prime example of this
phenomenon is provided by critical percolation: percolation
hulls intersect themselves at all scales, while the external
perimeters of percolation clusters are simple curves in the
universality class of self-avoiding random walks.

In the absence of boundaries in the system critical curves are
closed loops. If the system has a boundary and the boundary
condition changes at certain boundary points, it gives rise to
critical curves which start and end at these points. A hole in
the system is a component of the boundary. If the boundary
condition changes on it, we will observe critical curves
emanating from the hole. On scales much larger than the size of
the hole the latter becomes a puncture. In this case there are
critical curves coming out of a single point in the bulk of the
system. Inserting a puncture at some point in the bulk we can
ensure that in each realization of the statistical ensemble
there is a ``star'' --- a fixed number $k$ of curves starting
from this point.

\label{star page} Generally speaking, a critical system need
not contain punctures or be restricted in any way by insertion
of curve-creating operators. These are merely artificial
devices which we will use in our calculations. The meaning of
$k=2$ stars is obvious: any point on a curve is such a star. In
fact, stars with $k > 2$ can also be meaningful even in the
absence of punctures. A star with an even number $k \geqslant
4$ of legs is a point of self-intersection of a critical curve.
Since in the dilute phase the latter are simple curves with
probability one, stars with $k \geqslant 4$ can only appear as
exceptional configurations in rare realizations of the
statistical ensemble, all such realizations having the total
weight (probability) zero. In fact, we can quantify this
discussion by assigning a negative fractal dimension $d_k$ (see
Eq. (\ref{d_k}) below) to the set of stars with $k \geqslant 4$
legs. The meaning of a negative fractal dimension is as
follows. If in a system of size $L$ several curves approach
each other within a distance $l$, then on the scales much
larger than $l$ the configuration of the curves will look like
a star. However, in a typical realization of the statistical
ensemble, the number of points where such approach happens,
scales as $(L/l)^{d_k}$, and goes to zero in the thermodynamic
limit.

In the rest of the paper we apply methods of CFT to the problem
of the mixed multifractal spectrum of harmonic measure and
winding of critical curves. In this approach a $k$-legged star
of critical curves is created by an insertion of a specific
curve-creating operator $\psi_{0,k/2}$ \cite{Bauer-Bernard-CMP,
BRGW-PRL, RBGW-long-paper}. This operator can be realized as a
vertex operator in the Coulomb gas formulation of CFT, and has
known holomorphic charge $\alpha_{0,k/2}$ and weight
$h_{0,k/2}$, see Eqs. (\ref{central charge}, \ref{CC-charge},
\ref{CC-weights}) below. Other primary operators inserted close
to the curve-creating one can then serve as probes for the
harmonic measure. This has been explained and used in our
previous works \cite{BRGW-PRL, RBGW-long-paper}. In the present
paper we will see that to account for winding of critical
curves, the probes must have complex charges and conformal
weights.

We want to point out here that while harmonic measure can
behave nontrivially and be probed both in the bulk and on a
boundary of a critical system, the mixed spectrum that takes
into account winding is only defined in the bulk. Indeed, there
is no way a critical curve can rotate around a point on a
boundary. Therefore, our present analysis only deals with the
bulk curve creating operators.

\section{The case of a single point}
\label{one point}

For simplicity we start with the case of the electric field
(both the magnitude and the direction) measured at a single
point and then generalize the method to multiple points.

\subsection{Definition of multifractal exponents}
\label{definition of exponents}

Consider a $k$-legged star --- $k$ critical curves emanating
from the origin in the bulk of the system. We are interested in
the properties of the star much closer to the origin than the
system boundaries. The shape of the legs of the star far from
the origin is unimportant, which allows us to trace each leg
only up to a certain distance from the origin. Let us describe
the shape of such a truncated star by an analytic function
$w(z)$ which maps the exterior $\mathbb{S}$ of the star onto
the exterior of the unit disk $\mathbb{C}\backslash\mathbb{D}$.
The statistical ensemble of the shapes of the star defines the
statistical ensemble of $w(z)$.
\begin{figure}[t]
\centering
\includegraphics[width=0.8\textwidth]{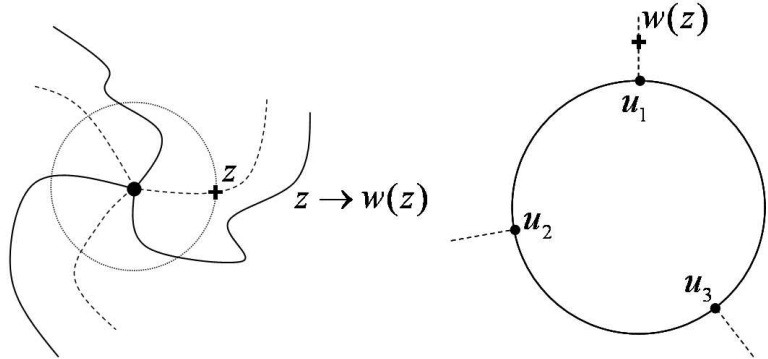}
\caption{The conformal transformation blows up the star into
the shape of a disk, like a puffer fish. The origin of the star
is mapped to several points $u_i$. The point $z$ is chosen to lie on
the dotted circle of a fixed radius and so that $w(z)$ lies on
the same radial line as $u_1$.}
\label{fig1}
\end{figure}
Close to the origin the $k$ curves of the star divide the plane
into $k$ sectors. Approached from different sectors, the origin
is mapped onto $k$ points $u_i$ on the unit circle: $\lim_{z\to
0} w(z) = u_i$ for $z$ in the $i$-th sector. To make $w(z)$
unique we specify $w(\infty)=\infty$ and $w'(\infty) =
\rho^{-1} > 0$, where $\rho$ is the conformal radius of the
star. Its fluctuations are unimportant. For each realization of
the star we can choose the truncation radius so that $\rho=1$.

As we have explained above, the magnitude and the direction of
the electric field at a point $z$ are determined by $|w'(z)|$
and $\arg w'(z)$ respectively. Both these quantities are
random. Conformal invariance suggests that locally $|w'(z)|$
scales as a power of $|z|$ and $\arg w'(z)$ behaves as $\lambda
\ln |z|$. Therefore, we can study the joint moments of
$|w'(z)|$ and $e^{\arg w'(z)}$ expecting them to scale as
\begin{align}
\mathbf{E}\big[|w'(z)|^n e^{p\arg w'(z)}\big] &\sim
|z|^{\Delta_k(n,p)},
\label{definition}
\end{align}
where $\mathbf{E}$ denotes average over the fluctuating
geometry of the star. This equation defines the multifractal
exponents $\Delta_k(n,p)$, and our goal below is to obtain them
using methods of CFT.

Let us now discuss a subtle point about Eq. (\ref{definition}).
We can treat the average over the star shapes in two ways. The
first way is to specify the position of point $z$ in such a way
that for any star $\mathbb{S}$ this point is on the slanted
Green line $\gamma_\theta$ in, say, the first sector (the
choice of the sector is arbitrary and immaterial). Let us
denote this choice by $z_\theta$ (implicitly keeping in mind
that the choice depends on the star shape). Then the average in
the left hand side of Eq. (\ref{definition}) should be
understood as
\begin{align}
\mathbf{E}\big[|w'(z_\theta)|^n e^{p\arg w'(z_\theta)} \big]
&\sim |z_\theta|^{\Delta_k(n,p)},
\label{fixed theta}
\end{align}
(only the argument of $z_\theta$ fluctuates here, while
$|z_\theta|$ is fixed). The simplest such choice is to place
$z$ on the electric field line $\gamma_{\pi/2}$. Then in the
$w$ plane the image $w(z)$ lies on a radial straight line:
\begin{align}
\arg w(z_{\pi/2}) = \arg u_1,
\label{condition1}
\end{align}
see Fig. \ref{fig1} for illustration.

On the other hand, as we have argued in the previous section,
the scaling of $|w'(z)|$ and $e^{\arg w'(z)}$ should be the
same along {\it any} line $\gamma_\theta$. Therefore, we can
average (integrate) over $\theta$ (and sum over the different
sectors) in Eq. (\ref{fixed theta}) while keeping $|z_\theta|$
constant. This procedure produces a manifestly rotationally
invariant average. In this average we can distribute the
contributions that come from star shapes which differ by a
global rotation around the origin into sets, and then sum up
all such sets. Due to isotropy of our critical system, all
terms in a given set have the same statistical weight.
Therefore, to calculate the average, we can choose any single
term from each set and sum these up. We can choose these terms
so that $\arg z$ is the same for all of them, which gives the
left hand side in Eq. (\ref{definition}) with a {\it fixed}
point $z$. This is the second way to interpret the average over
the star shapes. As we see, the two ways are equivalent and
produce the same scaling behavior with the multifractal
exponents $\Delta_k(n,p)$.

\subsection{CFT calculation of multifractal exponents}

A critical system is characterized by the central charge $c$
which we parametrize, as is usual in CFT, by the so called
background charge $\alpha_0$:
\begin{align}
c &= 1 - 24\alpha_0^2.
\label{central charge}
\end{align}
As we have shown in Ref.  \cite{RBGW-long-paper}, in the
Coulomb gas description of the dilute phase we have to impose
the Dirichlet conditions on the system boundaries. A consistent
way of doing it also implies that the background charge
$\alpha_0$ in Eq. (\ref{central charge}) is non-positive:
$\alpha_0 \leqslant 0$.

Within the CFT description (see \cite{RBGW-long-paper, cft} for
introduction to the mathematical apparatus of CFT used below) a
$k$-legged star centered at the origin is created by inserting
the curve-creating operator $\psi_{0,k/2}(0)$ \cite{BRGW-PRL,
RBGW-long-paper}. This operator has holomorphic charge
\begin{align}
\alpha_{0,k/2}=\alpha_0 -\frac{k}{4}\big(\alpha_0 -
\textstyle\sqrt{\alpha_0^2 + 1} \big),
\label{CC-charge}
\end{align}
and its antiholomorphic charge is $\bar \alpha_{0,k/2} = -
\alpha_{0,k/2}$. It is spinless with real conformal weights:
\begin{align}
h = \bar h = h_{0,k/2} =
\alpha_{0,k/2} (\alpha_{0,k/2} - 2\alpha_0).
\label{CC-weights}
\end{align}

To calculate the mixed multifractal exponents $\Delta_k(n,p)$
by means of CFT we consider a primary operator $\cO_{h, \bar
h}$ whose holomorphic and antiholomorphic weights $h, \bar h$
are complex and conjugate to each other:
\begin{align}
\bar h=h^*.
\end{align}
(Notice that we use asterisk to denote complex conjugation,
while the bar over a weight or charge simply indicates the
antiholomorphic sector.) Its holomorphic and antiholomorphic
charges are related to the weights by the equations
\cite{RBGW-long-paper}
\begin{align}
h&=\alpha(\alpha-2\alpha_0), & \bar h &= \bar\alpha(\bar\alpha
+ 2\alpha_0).
\end{align}
(The introduction of the charges is convenient because they
make fusion of primary operators look simple: the charges of
the fused fields simply add up.) Note the sign difference in
the formula for $\bar h$ compared to that in Ref.~\cite{cft}.
The difference is necessitated by our convention ($\alpha_0
\leqslant 0$) for the dilute phase. The solutions $\alpha(h)$,
$\bar\alpha(\bar h)$ are chosen so that
$\alpha(0)=\bar\alpha(0)=0$:
\begin{align}
\alpha &= \alpha_0 + \scriptstyle{\sqrt{\textstyle \alpha_0^2 + h}}, &
\bar\alpha &= -\alpha_0 - \scriptstyle{\sqrt{\textstyle
\alpha_0^2 + \bar h}}.
\end{align}
The two charges are related by $\bar\alpha = -\alpha^*$. If we
denote the real and imaginary parts of the holomorphic charge
as $\alpha'$ and $\alpha''$, the real and imaginary parts of
the holomorphic weight $h$ can be written as
\begin{align}
\text{Re} \, h &= \alpha'^2 - \alpha''^2 - 2\alpha_0\alpha', &
\text{Im} \, h &= 2\alpha''(\alpha' - \alpha_0).
\label{Re-Im-h}
\end{align}

The operator $\cO_{h, \bar h}$ plays the role of a ``probe'' of
harmonic measure and rotation of critical curves when we place
it at a point $z$ near the origin of a $k$-star. The star is
created by inserting the curve-creating operator
$\psi_{0,k/2}(0)$. Thus we propose to consider the following
correlation function:
\begin{align}
\Big\langle \psi_{0,k/2}(0)\cO_{h,\bar h}(z)\Psi(\infty)\Big\rangle
\propto \big|z^{2\alpha_{0,k/2} \alpha}\big|^2
= |z|^{4\alpha_{0,k/2} \alpha'} e^{-4\alpha_{0,k/2} \alpha'' \arg z}.
\label{multivalued C}
\end{align}
Here and in what follows these angular brackets denote a CFT
average and $\Psi$ stands for all operators far from the origin
and is necessary to make the correlation function non-zero. The
right hand side in Eq. (\ref{multivalued C}) was obtained in a
standard way by fusion of primary operators (expressed as
vertex operators in the Coulomb gas formulation of CFT). Notice
that the presence of complex weights and charges leads to an
expression that is explicitly angular dependent and
multivalued. Thus, this correlator does not correspond to any
physical quantity, and we need to consider
\begin{align}
C = \Big\langle e^{4\alpha_{0,k/2}\alpha''\arg z}
\psi_{0,k/2}(0)\cO_{h,\bar h}(z)\Psi(\infty)\Big\rangle
\propto |z|^{4\alpha_{0,k/2}\alpha'}.
\label{C}
\end{align}
In this formulation the position of the point $z$ is fixed, and
the extra exponential factor is a constant that can be pulled
out of the average $\langle \ldots \rangle$.

The same correlation function $C$, being an average over all
degrees of freedom in the system, can be evaluated in another
way, using the so-called two-step averaging procedure
\cite{Bauer-Bernard-big-review, BRGW-PRL, RBGW-long-paper,
Bauer-Bernard-Houdayer}. We can first fix the shape of the star
and sum over the rest of the degrees of freedom. Then we
average over all possible shapes of the star:
\begin{align}
C = \mathbf{E}\Big[ e^{4\alpha_{0,k/2}\alpha''\arg z}
\big\langle \cO_{h,\bar
h}(z)\Psi(\infty)\big\rangle_{\mathbb{S}}\Big].
\label{C fixed z}
\end{align}
The subscript of a correlation function refers to the domain in
which it is computed, in this case the exterior of the star
$\mathbb{S}$.

The point $z$ is still fixed in Eq. (\ref{C fixed z}),
therefore, for each shape $\mathbb{S}$ this point lies on a
different slanted Green line $\gamma_\theta$. However, as in
Section \ref{definition of exponents}, we can argue that the
scaling with $|z|$ is the same (independent of $\theta$) for
all the terms with different values of $\theta$. Then to find
the exponents $\Delta_k(n,p)$ it is sufficient to retain only
terms with a fixed (in each star configuration) value of
$\theta$ and consider a different quantity
\begin{align}
C_\theta = \mathbf{E}\Big[ e^{4\alpha_{0,k/2}\alpha''\arg z_\theta}
\big\langle \cO_{h,\bar h}(z_\theta)\Psi(\infty)
\big\rangle_{\mathbb{S}}\Big].
\label{C fixed theta}
\end{align}
There is a parallel between the correlator $C$ and the
expression in Eq. (\ref{definition}), and the correlator
$C_\theta$ and the expression in Eq. (\ref{fixed theta}):
though $C$ and $C_\theta$ are formally very different objects,
they scale with $|z|$ in the same way. The crucial difference
between $C$ and $C_\theta$ is that in the latter object the
argument of $z_\theta$ is fluctuating (it depends on the star
configuration) and the factor $e^{4\alpha_{0,k/2}\alpha''\arg
z_\theta}$ must be kept under the average $\mathbf{E}[\ldots]$.

The next step in transforming the expression (\ref{C fixed
theta}) is to map the exterior of the star $\mathbb{S}$ onto
the exterior of the unit disk by $w(z)$. Under this map the
operator $\cO_{h,\bar h}$ transforms according to the
definition of a CFT primary field, and operators at infinity do
not transform due to the normalization of $w(z)$. Then
conformal covariance of correlators of CFT implies
\begin{align}
C_\theta =\mathbf{E}\Big[e^{4\alpha_{0,k/2}\alpha''\arg z_\theta}
w'(z_\theta)^h\bar w'(\bar z_\theta)^{\bar h} \big\langle\cO_{h,\bar
h}(w(z_\theta))\Psi(\infty)\big\rangle_{\mathbb{C}
\backslash\mathbb{D}}\Big].
\label{2}
\end{align}
In this correlator the value of the angle $\theta$ is
immaterial. For simplicity of notation let us choose it to be
$\theta = \pi/2$, which means that the point $z_{\pi/2}$ is
chosen according to Eq. (\ref{condition1}) and lies on the
unique electric field line connected to the origin of the star.
We will drop the subscript $\pi/2$ in the subsequent equations.

In the $w$ plane, $\arg w(z)$ is a fixed number no greater than
$2\pi$. At the same time, when $w(z)$ approaches $u_1$, its
pre-image $z$ moves along the electric field line and exhibits
a lot of winding. Since $\arg w(z) \sim \arg w'(z) + \arg z$,
we can write
\begin{align}
\arg z\sim -\arg w'(z),
\label{arg}
\end{align}
because both angles here are large. Thus, we can replace $\arg
z$ by $-\arg w'(z)$ in Eq. (\ref{2}). Using Eq. (\ref{Re-Im-h})
we can also rewrite
\begin{align}
w'(z)^h \bar w'(\bar z)^{\bar h} &= \big|w'(z)^h \big|^2 =
|w'(z)|^{2 \, \text{Re}\,h} e^{-2 \, \text{Im}\, h \arg w'(z)}
\nonumber \\
&= |w'(z)|^{2(\alpha'^2 - \alpha''^2 - 2\alpha_0\alpha')}
e^{- 4\alpha''(\alpha' - \alpha_0) \arg w'(z)}.
\end{align}

The remaining CFT correlation function in the exterior of the
disk is computed by fusion of the primary field with the
boundary (which means the fusion of the holomorphic part
$\cO_h(w(z))$ with its image across the boundary of the unit
disk). According to Eq. (\ref{condition1}), both $w(z)$ and its
image in the unit circle $1/w^*(z)$ have the same (constant)
argument $\arg u_1$. Then we find the following scaling
behavior:
\begin{align}
\big\langle \cO_{h,\bar h}(w(z))
\Psi(\infty)\big\rangle_{\mathbb{C}\backslash\mathbb{D}}
&\propto \Bigl(w(z)-\frac{1}{w^*(z)}\Bigr)^{-2\alpha\bar\alpha}
\sim |w(z)-u_1|^{2|\alpha|^2}
\nonumber \\
&\sim |z|^{2(\alpha'^2+\alpha''^2)} |w'(z)|^{2(\alpha'^2+\alpha''^2)}.
\label{fusew}
\end{align}

At this stage the correlator $C_\theta$ evaluated in two steps
becomes
\begin{align}
C_\theta &\sim |z|^{2(\alpha'^2+\alpha''^2)}
\mathbf{E} \Big[|w'(z)|^{4\alpha'(\alpha' - \alpha_0)}
e^{- 4\alpha''(\alpha' - \alpha_0 + \alpha_{0,k/2}) \arg w'(z)} \Big].
\end{align}
Since, as we have argued, both $C$ and $C_\theta$ should scale
in the same way with $|z|$, comparison with Eq. (\ref{C}) gives
\begin{align}
\mathbf{E} \Big[|w'(z)|^{4\alpha'(\alpha' - \alpha_0)}
e^{- 4\alpha''(\alpha' - \alpha_0 + \alpha_{0,k/2}) \arg w'(z)} \Big]
\sim |z|^{-2(\alpha'^2+\alpha''^2) + 4\alpha'\alpha_{0,k/2}}.
\end{align}

The last equation is identical to the definition
(\ref{definition}) provided we denote
\begin{gather}
n = 4\alpha'(\alpha' - \alpha_0), \qquad p =
-4\alpha''(\alpha'-\alpha_0 + \alpha_{0,k/2}),
\label{n-p} \\
\Delta_k(n,p) = -2(\alpha'^2+\alpha''^2) +
4\alpha'\alpha_{0,k/2}. \label{Delta-1}
\end{gather}
As usual, we need to choose the solutions of Eqs. (\ref{n-p})
that vanish for $n=0, p=0$:
\begin{align}
2\alpha' &= \alpha_0 + \textstyle\sqrt{\alpha_0^2 +n}, &
\alpha'' &= - \frac{p}{4(\alpha'-\alpha_0 + \alpha_{0,k/2})}.
\end{align}
Substituting these values into Eq. (\ref{Delta-1}) we find the
mixed multifractal exponents $\Delta_k(n,p)$. With the
shorthand
\begin{align}
\alpha_n = \alpha_0 + \textstyle\sqrt{\alpha_0^2 +n}
\end{align}
the answer can be written as
\begin{align}
\label{single}\Delta_k(n,p) = 2\alpha_{0,k/2}\alpha_n
- \frac{1}{2}\alpha_n^2 -
\frac{1}{8}\frac{p^2}{\left(\frac{1}{2}\alpha_n -\alpha_0 +
\alpha_{0,k/2}\right)^2}.
\end{align}
We can also rewrite this result as
\begin{align}
\Delta_k(n,p)= \Delta_k(n) - \frac{1}{4}\frac{p^2}{\Delta_k(n) + n
+ 2(\alpha_{0,k/2}-\alpha_0)^2},
\label{single-1}
\end{align}
where
\begin{align}
\Delta_k(n) = \Delta_k(n,0) = 2\alpha_{0,k/2}\alpha_n
- \frac{1}{2}\alpha_n^2.
\label{Delta_k(n)}
\end{align}

\section{The general case}
\label{general case}

To define a multi-point generalization of the mixed spectrum
(\ref{definition}) we consider the following average:
\begin{align}
\label{definition2} \mathbf{E}\Bigl[e^{p\arg w'(z_1)}
\prod_{i=1}^k|w'(z_i)|^{n_i} \Bigr]\sim
r^{\Delta_k(\{n_i\},p)}.
\end{align}
Here all the points $z_i$ have the same distance to the star
origin: $|z_i|=r$, and no two of them lie in the same sector.
Because the curves do not intersect, all the winding angles
$\arg w'(z_i)$ differ by no more than $2\pi$, and since they
are all very large, they must all scale in the same way: $\arg
w'(z_i)\sim\arg w'(z_1)$. Thus, the topology of the star leads
to only one parameter $p$ describing the rotation, while we
have $k$ parameters $n_i$ for harmonic measure in each of the
sectors between the star's legs.

Similar to the discussion in Section \ref{definition of
exponents} the points $z_i$ can be either fixed or can be
chosen to lie on specific slanted Green lines. These choices do
not affect the scaling with $r$ in Eq. (\ref{definition2}). For
example, we can choose $\arg z_i$ by the requirement
\begin{align}
\arg w(z_i) = \arg u_i.
\label{condition2}
\end{align}
Then in all realizations of the star each $z_i$ lies on the
unique Green line in the $i$-th sector that starts at the
star's origin.

The calculation of $\Delta_k(\{n_i\},p)$ is a straightforward
generalization of the calculation of the previous section. We
introduce $k$ primary operators $\cO_{h_i,\bar h_i}(z_i)$ each
with complex conjugate weights: $\bar h_i = h^*_i$ and
holomorphic charges $\alpha_i = \alpha'_i + i\alpha''_i$. In
close parallel to Eq. (\ref{C}) we consider a CFT correlation
function:
\begin{align}
C_k &= \Bigl\langle e^{4\alpha_{0,k/2}\sum_i\alpha_i''\arg z_i +
4\sum_{i < j} (\alpha'_i\alpha''_j + \alpha''_i \alpha'_j) \arg (z_i-z_j)}
\nonumber\\
& \quad \times \psi_{0,k/2}(0) \prod^k_{i=1} \cO_{h_i,\bar
h_i}(z_i)\Psi(\infty) \Bigr\rangle.
\end{align}
This correlation function can be computed by fusion of primary
fields:
\begin{align}
&\Bigl\langle \psi_{0,k/2}(0) \prod^k_{i=1} \cO_{h_i,\bar
h_i}(z_i)\Psi(\infty) \Bigr\rangle \propto
\prod_i \big|z_i^{2\alpha_{0,k/2} \alpha_i} \big|^2
\prod_{i<j} \big|(z_i - z_j)^{2\alpha_i \alpha_j} \big|^2 \nonumber \\
&= \prod_i |z_i|^{4\alpha_{0,k/2} \alpha'_i}
e^{-4\alpha_{0,k/2} \alpha''_i \arg z_i} \nonumber \\
& \quad \times
\prod_{i<j} |z_i - z_j|^{4(\alpha'_i \alpha'_j - \alpha''_i \alpha''_j)}
e^{-4(\alpha'_i\alpha''_j + \alpha''_i \alpha'_j) \arg (z_i-z_j)}.
\end{align}
The exponential factors in the definition of $C_k$ are chosen
to cancel the angular dependence in the last equation. Since in
our setup $|z_i|= r$ and $|z_i-z_j| \sim r$ for all $i$ and
$j$, the correlation function $C_k$ scales as
\begin{align}
C_k \sim r^{4\alpha_{0,k/2}\sum_i\alpha_i'+ 4 \sum_{i<j}
(\alpha'_i\alpha'_j - \alpha''_i\alpha''_j)}.
\label{genz}
\end{align}

On the other hand, we can consider the similar correlator
$C_{k,\theta}$ where the points $z_i$ are chosen according to
Eq. (\ref{condition2}) and, therefore, fluctuate form
configuration to configuration. The correlators $C_k$ and
$C_{k,\theta}$ should scale as the same powers of $r$. We can
evaluate the correlation function $C_{k,\theta}$ in two steps:
first we fix the shape of the star and sum over the rest of the
degrees of freedom. Near a fixed star all the arguments $\arg
z_i$ and $\arg (z_i-z_j)$ are large and differ by no more than
$2\pi$. Therefore we can replace them all by one of them, say
$\arg z_1$. Then we average over all possible shapes of the
star:
\begin{align}
C_{k,\theta} \sim \mathbf{E} \Big[ e^{4\left
(\alpha_{0,k/2}\sum_i\alpha_i'' + \sum_{i\neq j}
\alpha'_i\alpha''_j \right)\arg z_1} \Big\langle \prod_i
\cO_{h_i,\bar h_i}(z_i)\Psi(\infty) \Big\rangle_{\mathbb{S}}\Big].
\end{align}
Now we map the exterior of the star $\mathbb{S}$ onto the
exterior of the disc $\mathbb{C}\backslash\mathbb{D}$ and note
that $\arg z_1 \sim -\arg w'(z_1)$ (compare with Eqs. (\ref{2},
\ref{arg})):
\begin{align}
C_{k,\theta} &\sim \mathbf{E} \Big[e^{-4\left
(\alpha_{0,k/2}\sum_i\alpha_i'' + \sum_{i\neq j}
\alpha'_i\alpha''_j \right)\arg w'(z_1)} \nonumber \\
&\quad \times \prod_i \big|w'(z_i)^{h_i}\big|^2
\Big\langle\prod_i\cO_{h_i,\bar h_i}(w(z_i))
\Psi(\infty)\Big\rangle_{\mathbb{C}\backslash\mathbb{D}}\Big].
\end{align}

The remaining correlator in $\mathbb{C}\backslash\mathbb{D}$ is
computed by fusion of primary fields with the boundary (compare
with Eq. (\ref{fusew})):
\begin{align}
\Big\langle\prod_i\cO_{h_i,\bar h_i}(w(z_i))
\Psi(\infty)\Big\rangle_{\mathbb{C}\backslash\mathbb{D}} \sim
\prod_i |w(z_i) - u_i|^{-2\alpha_i\bar\alpha_i} \sim \prod_i
\big(r |w'(z_i)| \big)^{2 |\alpha_i|^2}.
\end{align}
There are other factors in this correlator that are coming from
the fusion of different bulk fields $\cO_{h_i,\bar
h_i}(w(z_i))$ with each other. These cross-terms look like
$\big|(w(z_i) - w(z_j))^{2\alpha_i \alpha_j}\big|^2$, and do
not contribute to the short-distance behavior, since all $z_i$
lie in different sectors of the star, and the difference
$w(z_i) - w(z_j)$ stays finite in the limit $z_i \to 0$. As a
result,
\begin{align}
\label{genw} C_{k,\theta} \sim  r^{2\sum_i (\alpha'^2_i + \alpha''^2_i)}
\, \mathbf{E}\Big[e^{p\arg w'(z_1)}\prod_i
|w'(z_i)|^{n_i}\Big],
\end{align}
where
\begin{align}
p &= -4 \sum_i\alpha''_i \Big(\sum_j\alpha'_j + \alpha_{0,k/2} -
\alpha_0\Big), &
n_i &= 4\alpha'_i(\alpha'_i - \alpha_0).
\end{align}
We solve these equations for $\alpha'_i$ and $\sum_i\alpha''_i$
and choose the solutions which vanish at $n_i = p =0$. Then
Eqs. (\ref{genz}, \ref{genw}) imply Eq. (\ref{definition2})
with
\begin{align}
\Delta_k(\{n_i\},p) &= \sum_i \big(
2\alpha_{0,k/2}\alpha_{n_i} - \alpha_{n_i}^2 \big)
+ \frac{1}{2} \Big( \sum_i \alpha_{n_i} \Big)^2 \nonumber \\
&\quad - \frac{1}{8}\frac{p^2}{\left(\frac{1}{2}\sum_i\alpha_{n_i}
-\alpha_0 + \alpha_{0,k/2}\right)^2},
\label{multiple}
\end{align}
where $\alpha_{n_i} = \alpha_0 + \sqrt{\alpha_0^2 +n_i}$ .
Similarly to Eq. (\ref{single-1}), this result can be rewritten
as
\begin{align}
\Delta_k(\{n_i\},p) &= \Delta_k(\{n_i\}) - \frac{1}{4}
\frac{p^2}{\Delta_k(\{n_i\}) + \sum_i n_i +
2(\alpha_{0,k/2}-\alpha_0)^2}, \label{multiple-1}
\\
\Delta_k(\{n_i\}) &= \sum_i \big(
2\alpha_{0,k/2}\alpha_{n_i} - \alpha_{n_i}^2 \big)
+ \frac{1}{2} \Big( \sum_i \alpha_{n_i} \Big)^2.
\end{align}
If all $n_i$ vanish except one, the formulas (\ref{multiple},
\ref{multiple-1}) reduce to (\ref{single}, \ref{single-1}).

\section{Discussion}
\label{discussion}

The exponents $\Delta_k(n,p)$ and $\Delta_k(\{n\},p)$ that we
have obtained in Eqs. (\ref{single}, \ref{multiple}) are most
natural from the point of view of CFT and correlation functions
of primary operators. They are easily related to other
multifractal exponents defined directly in terms of harmonic
measure and winding. To exhibit this relation we start with a
special case.

Consider a closed critical curve. Let us cover it by circles of
radius $r$ with centers at $\zeta_i$. Let us denote the
harmonic measure within each circle by $H(\zeta_i,r)$, and the
winding angle of the Green line ending at $\zeta_i$ at distance
$r$ away from $\zeta_i$ by $\phi(\zeta_i,r)$. Then we can
define a sort of ``global'' mixed spectrum $\tau(n,p)$ by (see
Eq. (2) in Ref. \cite{Duplantier-Binder})
\begin{align}
\mathbf{E}\Big[\sum_i H^n(\zeta_i,r) e^{p \phi(\zeta_i,r)} \Big]
\propto r^{\tau(n,p)}.
\label{definition-tau}
\end{align}
If we set $n=p=0$ in the above equation, we simply get the
number of circles of radius $r$ needed to cover the curve. This
number should scale as $r^{-d_f}$, where $d_f$ is the fractal
dimension of the critical curve. Therefore, $ \tau(0,0) = -
d_f$.

It is natural to assume some sort of ergodicity, so that the
average of the sum in Eq. (\ref{definition-tau}) can be
replaced by the local average $\mathbf{E} [H^n e^{p\phi}]$
multiplied by the number of terms in the sum. If the local
average scales as $r^{\tilde \tau(n,p)}$, then the global and
the local exponents are related by
\begin{align}
\tau(n,p) = \tilde \tau(n,p) - d_f.
\end{align}

The harmonic measure $H(\zeta_i,r)$ by electrostatic analogy
should locally scale as $r |w'(z_i)|$, where $z_i$ is a point
that lies not on the curve at distance $r$ away from $\zeta_i$,
compare with Fig. \ref{fig1}. In the situation we describe here
each point on the curve can be viewed as the origin of a
2-legged star. Then we can relate our exponent $\Delta_2(n,p)$
to the local and the global exponents $\tilde \tau(n,p)$,
$\tau(n,p)$:
\begin{align}
\tilde \tau(n,p) &= \Delta_2(n,p) + n, &
\tau(n,p) &= \Delta_2(n,p) + n - d_f.
\end{align}
The fractal dimension $d_f$ of the curve is related to the
dimension of the curve creating operator $\psi_{0,1}$
\cite{bb}:
\begin{align}
d_f = 2 - 2h_{0,1}.
\label{d_f}
\end{align}
Finally, we get the following relation
\begin{align}
\tau(n,p) &= \Delta_2(n,p) + n + 2h_{0,1} - 2.
\end{align}
Substituting here the expressions (\ref{CC-weights},
\ref{single-1}) with $k=2$ we get
\begin{align}
\tau(n,p) &= \tau(n) - \frac{1}{4} \frac{p^2}{\tau(n) + b},
\label{tau-n-p}
\end{align}
where
\begin{align}
\tau(n) &= \Delta_2(n) + n + 2h_{0,1} - 2
= \frac{1}{2} \big(\textstyle\sqrt{\alpha_0^2 + n} +
\textstyle\sqrt{\alpha_0^2 + 1} \big)^2 - 2(\alpha_0^2 + 1)
\nonumber \\ & = \frac{n-1}{2} + \frac{25 - c}{24}
\bigg(\sqrt{\frac{24n + 1 - c}{25 - c}} - 1\bigg) \label{tau-n}
\end{align}
is the multifractal spectrum of the harmonic measure (see Eq.
(6.32) in Ref. \cite{Duplantier}), and
\begin{align}
b &= 2 (\alpha_{0,1} - \alpha_0)^2 + 2 - 2h_{0,1}
= 2(\alpha_0^2 + 1) = \frac{25 - c}{12}.
\end{align}
Formula (\ref{tau-n-p}) is exactly the same as Eq. (14) for
$\tau(n,p)$ in Ref. \cite{Duplantier-Binder}.

Generalized exponents $\tau_k(\{n_i\},p)$ can be defined
similarly to Eq. (\ref{definition-tau}):
\begin{align}
\mathbf{E}\Big[\sum_{k-\text{stars}} e^{p \phi(z_1,r)} \prod_{i=1}^k
H^{n_i}(z_i,r) \Big] \propto r^{\tau_k(\{n_i\},p)}.
\label{definition-tau-n}
\end{align}
In this definition the points $z_i$ are the same as in Eq.
(\ref{definition2}), and $H(z_i,r)$ denotes harmonic measure on
the portion of the star inside the $i$-th sector up to radius
$r$. As we have argued in Section \ref{formulation of problem},
$k$-legged stars with $k \geqslant 4$ may appear spontaneously
in a critical system, even though in a subset of the
realization of the statistical ensemble with the total measure
zero. The origins of the $k$-legged stars have a certain
fractal dimension $d_k$. In full analogy with Eq. (\ref{d_f}),
the fractal dimension $d_k$ is related to the scaling dimension
$2h_{0,k/2}$ of the operator $\psi_{0,k/2}$:
\begin{align}
d_k = 2- 2h_{0,k/2}.
\label{d_k}
\end{align}
Notice that, consistent with the discussion in Section
\ref{formulation of problem}, $d_k$ is negative for $k
\geqslant 4$.

Assuming ergodicity and the scaling $H(z_i,r) \sim r
|w'(z_i)|$, as before, we can relate the generalized exponents
$\tau_k(\{n_i\},p)$ to $\Delta_k(\{n_i\},p)$:
\begin{align}
\tau_k(\{n_i\},p) &= \Delta_k(\{n_i\},p) + \sum_i n_i + 2h_{0,k/2} - 2.
\end{align}
Using Eqs. (\ref{CC-weights}, \ref{multiple-1}) we can express
this result as
\begin{align}
\tau_k(\{n_i\},p) &= \tau_k(\{n_i\}) - \frac{1}{4}
\frac{p^2}{\tau_k(\{n_i\}) + b},
\label{tau-many-n-p}
\end{align}
where
\begin{align}
\tau_k(\{n_i\}) &= \Delta_k(\{n_i\}) + \sum_i n_i + 2h_{0,k/2} - 2
\nonumber \\
&= \frac{1}{2} \Big(\frac{k}{2}\big(\alpha_0 +
\textstyle\sqrt{\alpha_0^2 + 1} \big) + \sum_i
\textstyle\sqrt{\alpha_0^2 + n_i}\Big)^2 - 2(\alpha_0^2 + 1)
\label{tau-many-n}
\end{align}
are the ``higher multifractal exponents'' of Ref.
\cite{Duplantier}, Section 7.3.1. (To convince oneself of the
equivalence, one needs to substitute $2\alpha_0 =
\gamma(1-\gamma)^{-1/2}$, where $\gamma \leqslant 0$ is the
so-called string susceptibility exponent.)

Expressions (\ref{tau-n-p}, \ref{tau-n}, \ref{tau-many-n-p},
\ref{tau-many-n}) and their meaning and consequences are
analyzed in detail in Ref. \cite{Duplantier}. Here we only
mention that the Legendre transforms of these exponents (the so
called singularity spectra) have a direct geometrical meaning
of fractal dimensions of subsets of points with a given local
scaling of harmonic measure and winding.

In conclusion, we have shown how a detailed description of
stochastic geometry of critical conformally-invariant curves in
terms of their harmonic measure and winding can be obtained by
means of CFT, from correlation functions of certain primary
operators. The inclusion of winding (rotations) makes it
necessary to consider operators with complex weights.

\section{Acknowledgements}

This research was supported in part by NSF MRSEC Program under
DMR-0213745, NSF Career Award under DMR-0448820, NSF grant
under DMR-0645461 (AB), and the U.S. Department of Energy grant
DE-FG02-ER46311 (IR). We wish to acknowledge helpful
discussions with I. Binder and P. Wiegmann. While preparing
this paper for publication, we have learned about a related
work by B. Duplantier and I. Binder.

\end{document}